\documentclass{article}       
\usepackage{graphicx}
\begin{document}

\newtheorem{theorem}{Theorem}[section]
\newtheorem{lemma}[theorem]{Lemma}
\newtheorem{proposition}[theorem]{Proposition}
\newtheorem{corollary}[theorem]{Corollary}

\newenvironment{proof}[1][Proof]{\begin{trivlist}
\item[\hskip \labelsep {\bfseries #1}]}{\end{trivlist}}
\newenvironment{definition}[1][Definition]{\begin{trivlist}
\item[\hskip \labelsep {\bfseries #1}]}{\end{trivlist}}
\newenvironment{example}[1][Example]{\begin{trivlist}
\item[\hskip \labelsep {\bfseries #1}]}{\end{trivlist}}
\newenvironment{remark}[1][Remark]{\begin{trivlist}
\item[\hskip \labelsep {\bfseries #1}]}{\end{trivlist}}

\newcommand{\qed}{\nobreak \ifvmode \relax \else
      \ifdim\lastskip<1.5em \hskip-\lastskip
      \hskip1.5em plus0em minus0.5em \fi \nobreak
      \vrule height0.75em width0.5em depth0.25em\fi}

\title{Multi-Stratum Networks:\\toward a unified model of on-line identities
}

\author{Matteo Magnani\footnote{Data Intensive Systems Lab, Department of Computer Science, University of Aarhus, \r{A}bogade 34, 8200 Aarhus-N, Denmark. magnanim@cs.au.dk}  \and Luca Rossi\footnote{Department of Social Sciences, University of Urbino Carlo Bo, via Saffi 15, 61029 Urbino. luca.rossi@uniurb.it}}


\maketitle

\begin{abstract}
One of the reasons behind the success of Social Network Analysis is its simple and general graph model made of nodes (representing individuals) and ties. However, when we focus on our daily on-line experience we must confront a more complex scenario: people inhabitate several on-line spaces interacting to several communities active on various technological infrastructures like Twitter, Facebook, YouTube or FourSquare and with distinct social objectives. This constitutes a complex network of interconnected networks where users' identities are spread and where information propagates navigating through different communities and social platforms. In this article we introduce a model for this layered scenario that we call \emph{multi-stratum network}. Through a theoretical discussion and the analysis of real-world data we show how not only focusing on a single network may provide a very partial understanding of the role of its users, but also that considering all the networks separately may not reveal the information contained in the whole multi-stratum model.
\end{abstract}

\section{Introduction}

Researchers in Social Network Analysis (SNA) have developed methodologies and metrics that, although designed to describe and analyze communities of people, have been applied to several contexts ranging from epidemiology to computer networks. However, it has recently become clear that a thorough analysis of human communities may require models that can capture more than just the existence of generic connections between specific people. This need for more complex modeling tools appears from the proposal of multi-dimensional (also called multi-layer(ed) or multi-modal) network models, the emergence of research fields like Dynamic Network Analysis and the development of spatio-temporal Social Network models. While these models significantly increase our ability to study social phenomena, they are still populated by single individuals, often connected to each other with multiple kinds of ties, sometimes changing their location or behavior in time, but still represented using a construct as simple as a node in a network. However, while performing our social acts we cannot be considered indivisible units but, citing a novel by Luigi Pirandello, we are \emph{one, no one and one hundred thousand} at the same time \cite{Pepe11}.

To make this point more concrete, if we consider our contemporary social experience we can easily see that it is largely based on our ability to master a growing number of social contexts where we play different characters at the same time. Job interviews, chatting among friends or casual conversation at the bus stop, all these social interactions require a specific perception of our audience and the ability of choosing the right topic and tone among a list of many possibilities. We need to perform our social role starting from a clear definition of the social frame \cite{Goffman1974} we are performing in. 

When we move into the on-line context this scenario is, if possible, even harder to manage. On-line communication offers undoubtedly many opportunities to experience this multiplicity of identities and this has been one of the classical topics in Computer Mediated Communication studies \cite{Turkle1995}. Within this perspective on-line identity has often been described as a self conscious performance of identity practices \cite{boyd2010} played in different on-line contexts. The collapse of social contexts followed by the growing success of Social Network Sites (SNSs) \cite{Boyd2008} shifts this phenomenon even forward. Within SNSs like Facebook or even more prominently within microblogging sites like Twitter or Friendfeed users hardly have a clear perception of their final audience and the control of the social context where their communicative interactions are taking place is often a challenge \cite{boyd2010}. 

As a partial solution to this problem we saw over the last few years how, despite a general movement toward a reduction of the number of SNSs (largely due to the success of Facebook), dedicated social networks seem to be still very popular. Services like LinkedIn -- for professional networking -- or match.com -- for on-line dating -- show how, despite a huge social pressure pushing us toward a merging of our on-line presence, there is an interest in maintaining a higher control over specific areas of our on-line presence.
Therefore, at the moment, users' on-line presence is no longer scattered through hundreds of difference services but, at the same time, it is not yet unified within a single service (and it seems this will not happen anytime soon). 

On a metaphorical level contemporary on-line presence is similar to a pillar, representing a single user connected to other users on  several autonomous \emph{strata} (or \emph{floors} of this architecture made of multiple networks). Two users might be connected by many strata at the same time, e.g., two friends  may be \emph{friends} in Facebook, in Flickr and YouTube and \emph{followers} in Twitter -- while other users might be connected on just one stratum, e.g., co-workers connected only through LinkedIn.  Pillars are therefore the linkage between several networks each one representing only a stratum of the user's on-line presence.
The result is a highly complex system made of several strata, one for each on-line service where users can be present, and  where users (nodes) can be connected to each other through several edges (connections) with different properties. The way in which different users exploit the different kinds of available connections is related to their own personal strategy of on-line identity management. The choice about what to share and using which stratum/system is obviously part of this strategy. At the same time users have the ability to move conversations (or topics) from a stratum to another.

The impact of studying the whole system instead of each single network should be clear if we think of real and typical phenomena like on-line information diffusion. YouTube users are interconnected through \emph{following} and \emph{friendship} links allowing them to know when other users post new videos. However, news, \emph{memes} and almost any kind of on-line information usually spread through the network bouncing from a service to another, e.g.,  people post videos on YouTube but these videos often reach a high visibility when users start posting about them on their blogs, on Twitter or Facebook. This diffusion on Facebook may then have consequences on YouTube, where for instance new connections can be created between users who where exposed to their videos on Facebook. Studying this kind of phenomena in depth is impossible if we do not consider all the networks involved.

In this article we define what we call the Multi-Stratum Model (MSM). A MSM is able to deal with the high complexity of our contemporary on-line experience especially when we want to observe how information propagates through several networks at the same time or how on one side users act in different on-line social contexts dealing with different perceptions of their audience and -- at the same time -- their activity contributes to build their own on-line identity.

This paper, that extends both theoretically and experimentally our recent work on multi-layer networks \cite{DBLP:conf/asonam/MagnaniR11}, provides the following contributions:
\begin{itemize}
\item We define a model for the representation of multi-stratum networks.
\item We extend classical SNA centrality measures.
\item We report on the extraction and analysis of a real multi-stratum network.
\end{itemize}

The theoretical and practical analysis of MSMs presented in this article can be summarized in three main practical statements. First, not only studying a single network often limits our knowledge of on-line systems, but even studying separately the organization of multiple networks may not be sufficient to understand the overall role and position of some users. We call this phenomenon \emph{Network Gestaltism}, from the psychological theory stating that the whole may be greater than the sum of the parts. Second, by comparing SNA metrics computed on different networks we can understand how much the single networks are complementary to each other or have a similar social function. A \emph{Network Complementarity Index} quantifying this similarity can be very useful while building an \emph{efficient network portfolio} allowing the creation of an on-line identity without wasting resources, e.g., keeping several profiles on networks with similar social functions. Third, we show how the analyses performed on multi-stratum models can be influenced by the way in which data is collected, requiring special care during its application.

This article is organized as follows: in the next section we review existing work in this area. Then we formally define the multi-stratum model and extend centrality measures to apply to it. In Section~\ref{experiments} we describe the extraction and analysis of a real dataset with three strata using our model and metrics. We conclude with a discussion of the collected experimental evidence. 

\section{Related work}

This paper deals with at least three fields of related researches. On one side the MSM deals with the sociological researches on on-line presentation of the self by providing a formal model of the several \emph{strata} composing our everyday on-line presence. On  another side it deals with complex social network theories by offering a new approach for social network analysis and, at the same time, the final goal of the MSM is to provide a wider perspective on on-line propagation phenomena. Due to this highly interdisciplinary approach a large base of previous researches has to be taken into consideration.
 
On-line self presentation has to be understood as a technological mediated aspect of the broader issue of identity construction which can be described from a situationist perspective \cite{Garfinkel:1967uq} as an understanding of the context where the communicative interaction takes place leading to a specific choice about what kind of role is more suitable to perform \cite{Goffman1974}.

As clearly observed by Meyrowitz's \emph{No Sense of Place} \cite{Meyrowitz:1985fk} media change in a radical way the places where our communication happens. Electronic media -- from radio to television -- delete walls and boundaries that have often kept specific social contexts isolated. This phenomenon is not only affecting the way in which we define the distinction between what is public or what is private but affects also our possibility to have a clear perception of our \emph{audience} and therefore to have a clear understanding of the social context. The perception of the audience is therefore always more the perception of an \emph{imagined audience} which is used in order to perform our Self-conscious identity construction processes \cite{boyd2010}. This consciousness of an audience may change depending on what kind of on-line service we are studying \cite{Ellison2006} (in goal oriented spaces people are more conscious of that) nevertheless all on-line users have to be understood as multiple selections of possibilities operated according to the goals and the \emph{imaginated} audience of the on-line places they populate.

Research on complex social networks has often dealt with several kind of networks able to represent, within the same graph structure, different kinds of relationships between users \cite{DBLP:journals/tsmc/KazienkoMK11,DBLP:conf/socialcom/KazienkoBMG10} or even relationships between users with different attributes \cite{DBLP:conf/sigmod/ZhaoLXH11}.
These approaches have a different perspective than the approach that we are proposing, because they deal with the heterogeneity of ties and not the multiplicity of user identities. Nevertheless they point out several crucial aspects that have to be taken into careful consideration while we approach the description of large real-world social network sites. On one side the Multi-layered approach deals with the important aspect of users interacting in ways that cannot be described using a simple connection. Phenomena like users commenting on the same Facebook page while not directly connected through a \emph{friendship} connection or phenomena like the world-wide online conversations emerging through the use of Twitter's \emph{hashtags} \cite{mineros11} can be described with this kind of approach. At the same time the Multi-dimensional approach as it has been defined in \cite{DBLP:conf/sigmod/ZhaoLXH11} offers the opportunity to bring together the insights provided by the topological aspects of the network with the power of well established OLAP techniques.

As a part of the scenario of the most advanced applications of Social Network Analysis, Dynamic Network Analysis is an emerging research area partially related to the goal of our work. Dynamic Network Analysis brings together social network analysis, link analysis and multi-agent systems within network science. This interesting and emerging approach can describe networks with multiple kinds of nodes and multiple types of links connecting them \cite{Breiger:2003uq}, at the same time, similarly to the model we are proposing, nodes can belong to several networks at the same time. This approach and the related meta-matrix model \cite{Breiger:2003uq} have been widely adopted in organization studies \cite{Merrill:2006fk} and there has also been a great interest on the study of Dynamic Social Networks \cite{Kossinets:2006fk} that we expect to increase with respect to on-line networks in the future.

This multiplicity has to be taken into consideration when we start investigating the problem of on-line propagation. As we have highlighted in previous work \cite{MagnaniASONAM2010,MagnaniSCA2010} the choice between propagating or not a specific information item and through what kind of service doing that is mainly a choice related to our perception of our audience and related to what kind of identity we are constructing within that specific on-line place. In this paper we do not deal directly with information propagation, even though this is one of the main motivations of our model and will be object of further work, therefore we refer the reader to \cite{MagnaniBook2011} for more detailed references on factors enabling propagation in Social Network Sites, \cite{Keeling2005} for epidemiological models, \cite{Rushkoff1994} for its application to media, and \cite{Morley1980,Jenkins2010} about the active role of media audiences -- in addition to any text book on complex network analysis for the classical models. 

The contribution of our model in this context regards the fact that a theoretically well-founded research on propagation in SNSs should not try to simply show how information propagates through a SNS but also understand what is the role of SNS structures and connections in the larger process of propagation of cultural information. Within this process the belonging of every single node to several networks (that we define as social \emph{strata}) and the active role of every single node in acting through these strata according to his/her strategies or his/her perception of the audience should be part of every model aimed at understanding real on-line social phenomena.



\section{The multi-stratum model}

Our model extends the simpler network model used in traditional SNA using it as a building block of a more complex stratified architecture. A \emph{stratum} is therefore a simple graph. While several variants could be used here, like directed/undirected or weighted/unweighted graphs, in the following definitions we will use directed unweighted graphs. It is worth noticing again that this model is not an alternative to other models including aspects like weights, multi-dimensional ties or spatial annotations: all these orthogonal aspects can coexist inside the same model --- however in this paper we focus only on the coexistence of multiple networks/strata to avoid the introduction of unnecessary formalisms.

\begin{definition}[Stratum]
A Stratum is a graph $(V, E)$ where V is a set of vertexes and $E \subseteq V \times V$.
\end{definition}

\begin{figure}[htbp]
\begin{center}
\includegraphics[width=.33\columnwidth]{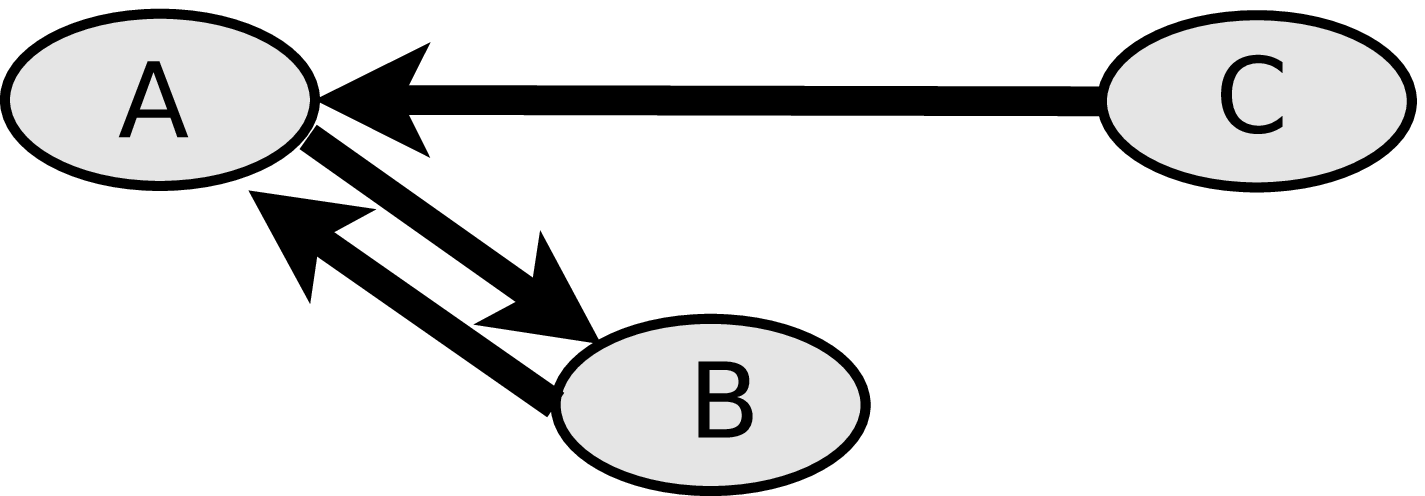}
\hspace{.5cm}
\includegraphics[width=.5\columnwidth]{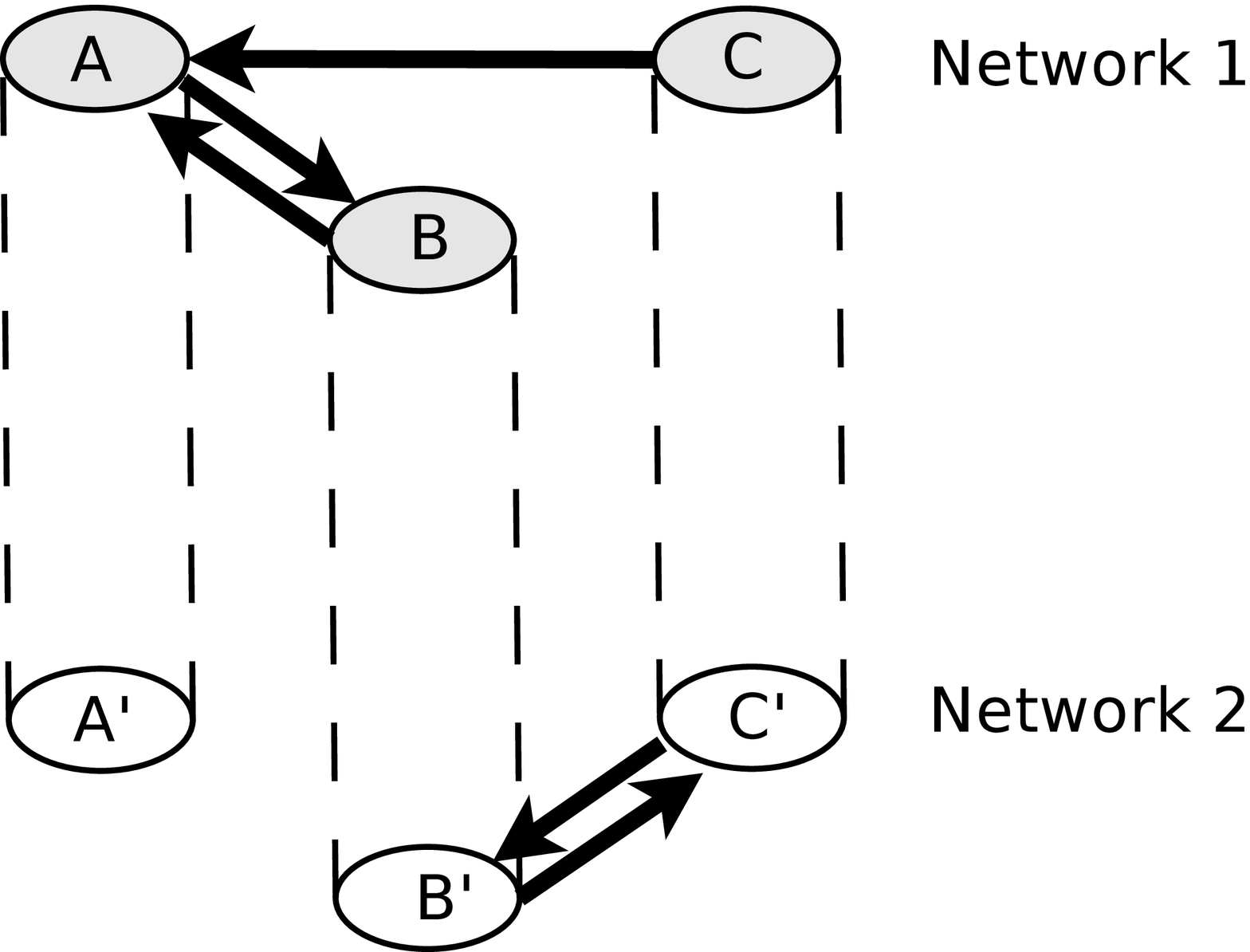}
\caption{A single-layer network (left) and a Pillar network (right).}
\label{model}
\end{center}
\end{figure}

\begin{example}
In the left hand side of Figure~\ref{model} we have represented a  single stratum with three users and three edges (without specifying the weights). This may correspond to a portion of a Twitter network, with following relationships.
\end{example}

When we start considering multiple networks, we need to know which nodes in one stratum correspond to nodes in the other. This is done using a Node Mapping.

\begin{definition}[Node Mapping]
A Node Mapping from a Stratum $L_1 = (V_1, E_1)$ to a Stratum $L_2 = (V_2, E_2)$ is a relation $m \subseteq V_1 \times V_2$ .
For each $u \in V_1$, the set $\mathcal{C}(u) = \{v \in V_2 \ |\ m(u,v) \}$ is the set of $V_2$ nodes corresponding to $u$.
\end{definition}

\begin{example}
The three users in the previous example may also have an account on Facebook, and here we can use another network to represent these three accounts and their relationships. This scenario has been represented in the right hand side of Figure~\ref{model} where we have exemplified how the connections between Facebook friends may not correspond to Twitter connections --- for instance a user does not necessarily follow the Twitter status of his friends, and often we follow public figures that are not our Facebook friends. In this example every user has exactly one account on each layer. For this reason we call this a \emph{Pillar model}, where every user can be seen as a pillar traversing several \emph{floors} and posed on the lower layer, not represented here, that is the level of physical reality, with a specific geographic location. A Pillar model is characterized by $|\mathcal{C}(u)| \in \{0,1\}$.
\end{example}

Finally we can define a general Multi Stratum Network, which consists of a set of networks and a matrix of Node Mappings.
\begin{definition}[Multi Stratum Network]
A Multi Stratum Network is a tuple $\mbox{MSN} = (L_1, \dots, L_n, \mbox{IM})$ where $L_i = (V_i,E_i)$ are Strata, IM (Identity Mapping\footnote{Here \emph{identity} should not be intended as a mathematical term, but as \emph{regarding the identity of the users}.}) is an $n \times n$ matrix of Node Mappings, with $\mbox{IM}_{i,j} \subseteq V_i \times V_j$, and MSN satisfies the following properties:
\begin{enumerate}
\item $\forall i, j : V_i \cap V_j = \emptyset$ (disjointness).
\item $\mbox{IM}_{i,i} = \{ (v, v) \ | \ v \in V_i \}$ (identity).
\item $(u, v) \in \mbox{IM}_{i,j} \iff (v, u) \in \mbox{IM}_{j,i}$ (symmetry).
\end{enumerate}
\end{definition}

\begin{figure}[htbp]
\begin{center}
\includegraphics[width=.7\columnwidth]{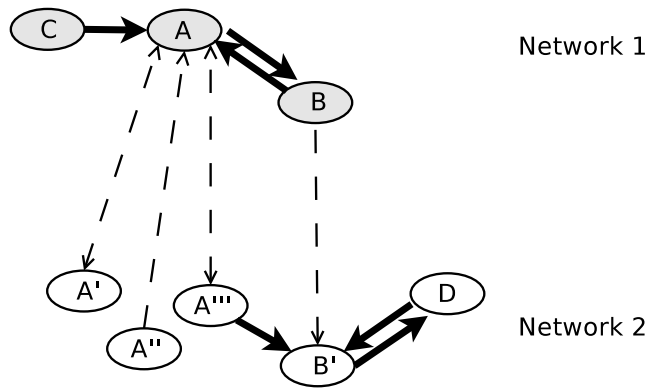}
\caption{A general Multi Stratum network}
\label{model1-N}
\end{center}
\end{figure}
 
\begin{example}
In Figure~\ref{model1-N} we have represented a multi stratum network more complex than a simple Pillar model. Here the same node in one network may correspond to multiple nodes in another. This is a typical case in social media aggregators, where for example we can have a \emph{BBC\_journalists} account following all registered BBC journalists and providing a single access point to all their updates. As a consequence, in this case nodes do not represent only users but more in general accounts.
\end{example}
 
\subsection{Analysis metrics}

In this section we extend basic SNA metrics to the context of Multi Stratum networks. We also motivate through several examples how these extensions may enhance our analysis power with respect to considering each network separately. Also in the following definitions we will not present all possible variations of these metrics (weighted/unweighted, etc.), but for the sake of clarity we will provide definitions corresponding to the Pillar model, that will be used in the subsequent experimental analysis.

\begin{figure}
\begin{center}
\includegraphics[width=.40\columnwidth ]{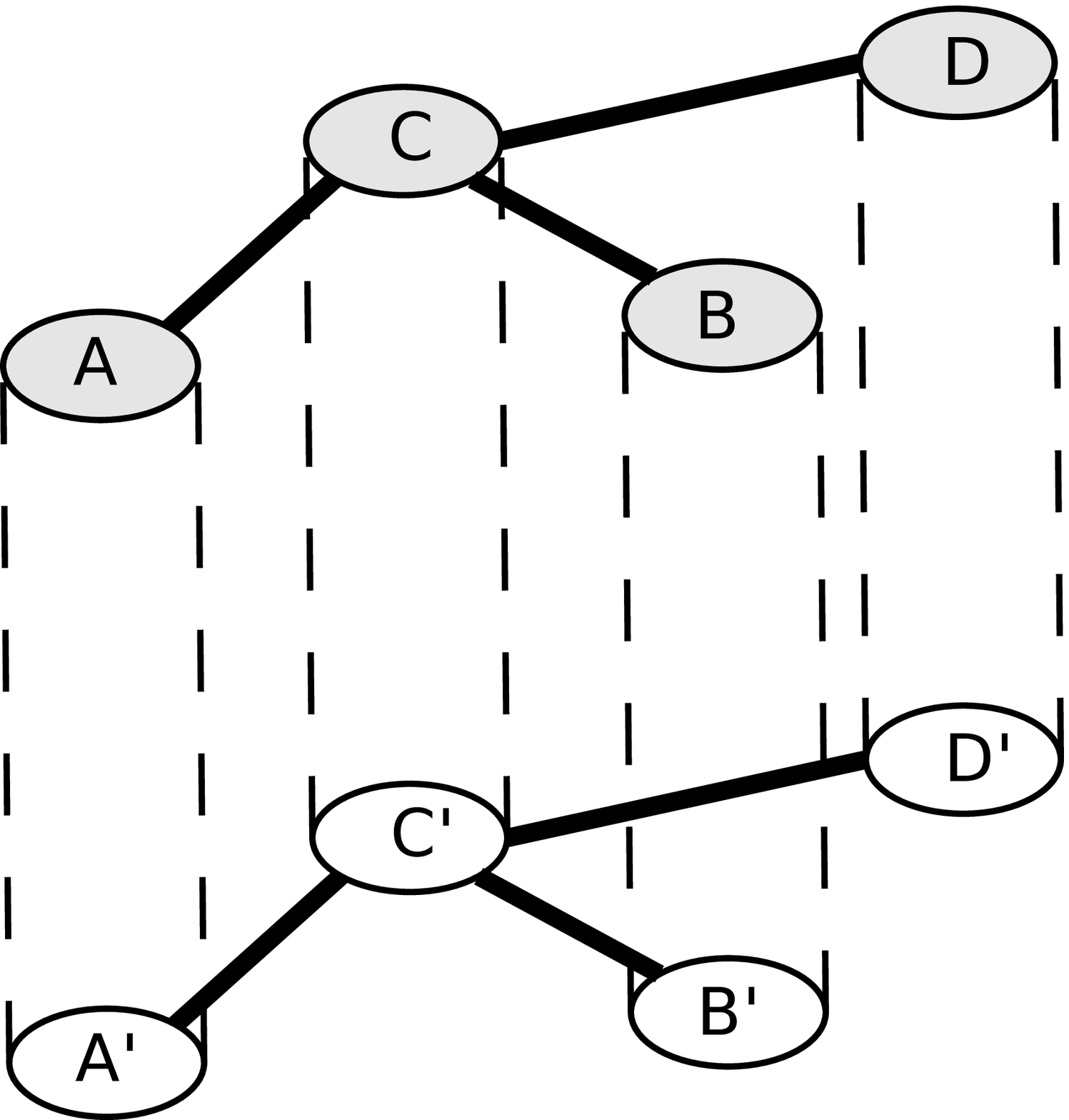} 
\hspace{.5cm}
\includegraphics[width=.47\textwidth ]{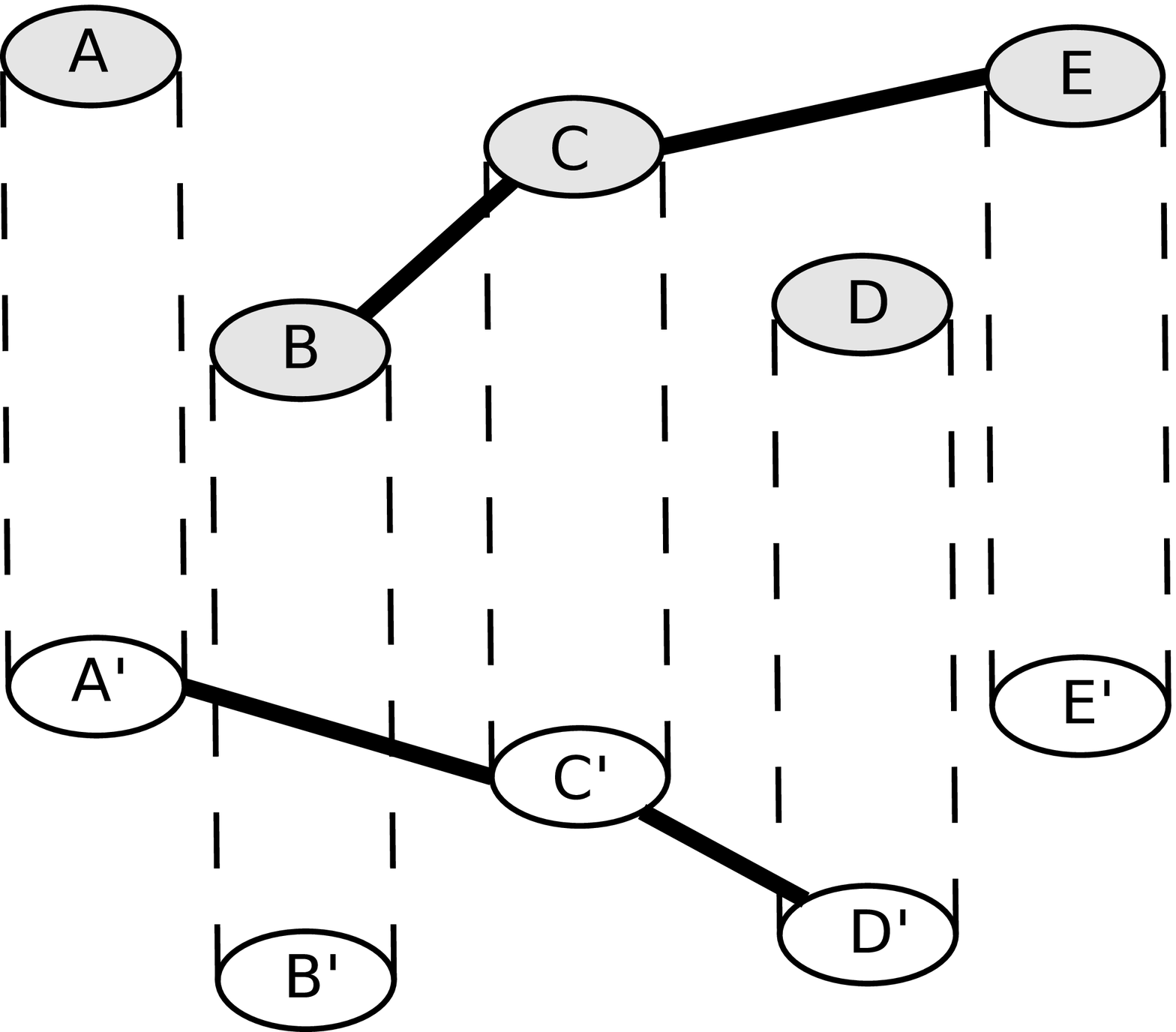} 
\end{center}
\caption{Multi Stratum degree centrality: a user (C) not exploiting the two networks to increase his/her audience (left) and a user increasing his/her audience (right)}
\label{degree_no}
\end{figure}

In the left hand side of Figure~\ref{degree_no} we have represented a user (C) with three connections in each network. If we look at the correspondences, however, we can see that the six connections include the same three users on both networks, therefore the Multi Stratum audience will be composed of six nodes but limited to three people. On the contrary, on the right we have represented a case where looking at the two single networks the central user would seem less connected than the one in the previous example. In fact he/she has two connections in the first network and two in the other, against the three and three of the previous example. However, this second user is exploiting the two networks in different ways, managing distinct audiences in one and the other. As a consequence, the overall degree centrality (measuring the number of \emph{real users} connected to C) will be 4, higher than in the previous example. 

The definition of the extended degree centrality just described by example is not immediate, and we can proceed in two steps. First, given a Multi Stratum Network $\mbox{MSN} = (L_1, \dots, L_n, \mbox{IM})$ and a node $v \in V_i$ we can reconstruct his/her identity using the identity mapping. More formally, we can consider an equivalence relation $eq$ where $eq(u, v) \iff \exists i, j : (u, v) \in \mbox{IM}_{i,j}$. This defines equivalence classes where each equivalence class represents the reconciliated identity of a user. As an example, in the left hand side of Figure~\ref{degree_no} $[A]_{eq} = \{A,A'\}$. Now we can build the following network starting from the MSN:

\begin{definition}[merge]
Let $\mbox{MSN} = (L_1, \dots, L_n, \mbox{IM})$ be a Multi Stratum Network. merge($MSN$) = $(V,E)$ where:
\begin{itemize}
\item $V = \{ [v]_{eq} \ | \ \exists i :  v \in V_i \}$
\item $E = \{ ([u]_{eq}, [v]_{eq}) \ | \  \exists i : (u,v) \in E_i \}$ 
\end{itemize}
\end{definition}

\begin{definition}[Multi Stratum Degree Centrality]
The Multi Stratum degree centrality of a user $u$ in a MSN is the Degree Centrality of $[u]_{eq}$ in merge(MSN).
\end{definition}

\begin{figure}
\begin{center}
\includegraphics[width=.40\columnwidth ]{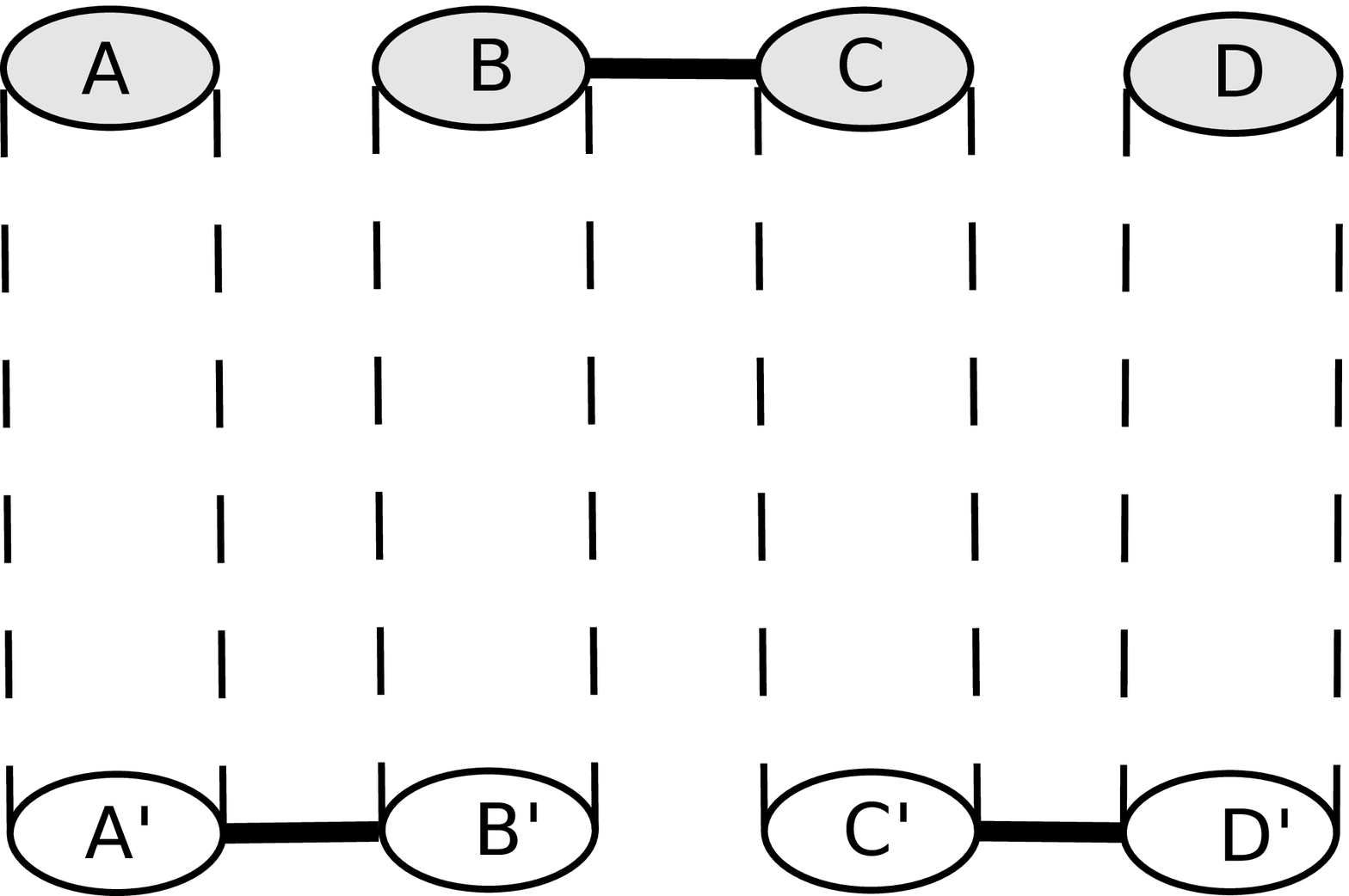} 
\hspace{.5cm}
\includegraphics[width=.53\textwidth ]{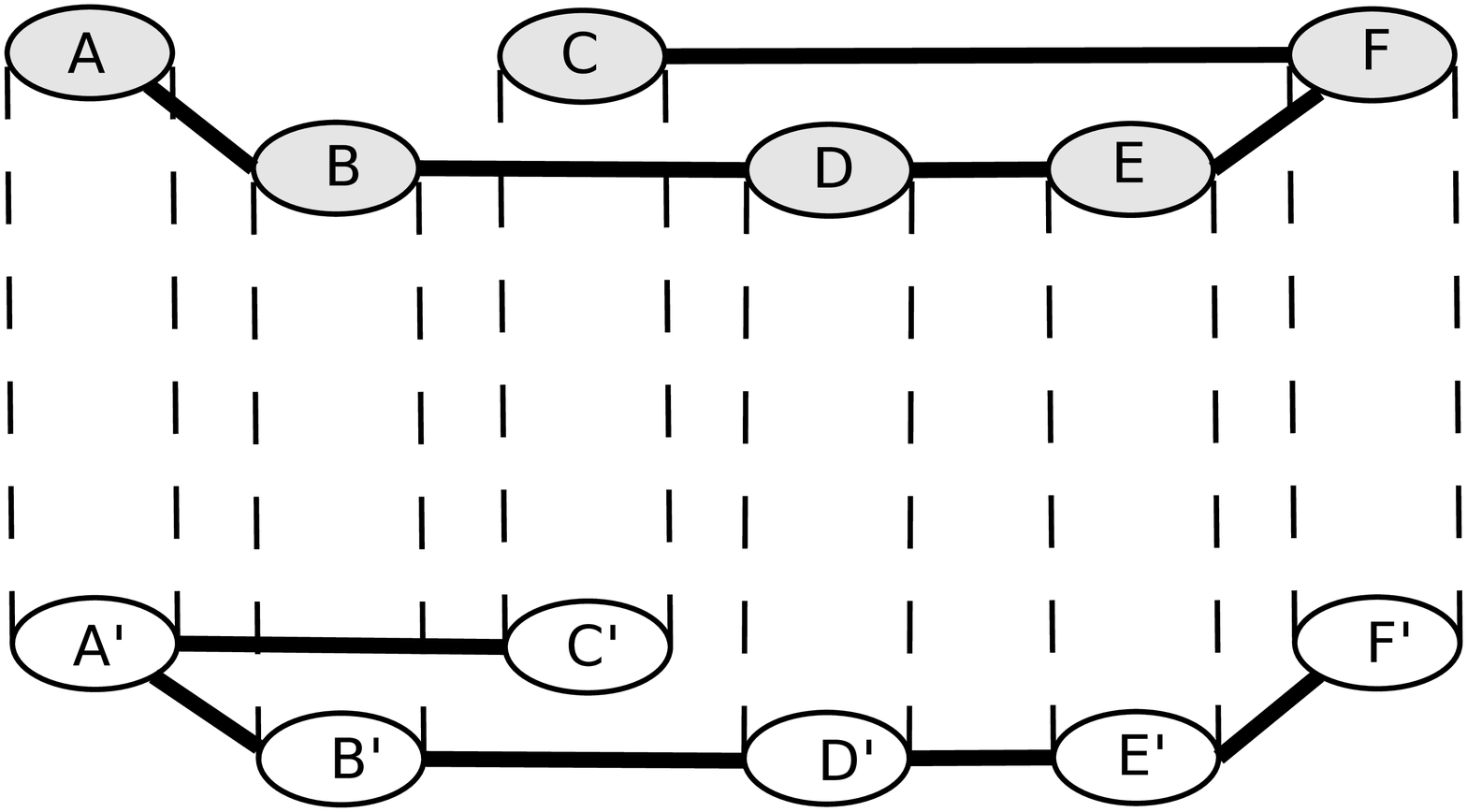} 
\end{center}
\caption{Node reachability (left) and node distances (right) in a Multi Stratum Network}
\label{reachability}
\end{figure}

Now we can consider two other basic metrics: closeness and betweenness.
The left hand side of Figure~\ref{reachability} shows that two users A and D whose accounts are not connected to each other can be in fact connected if we consider the Multi-Stratum network. In fact, an information item could reach D from A through the path A $\rightarrow$ A' $\rightarrow$ B' $\rightarrow$ B  $\rightarrow$ C $\rightarrow$ C' $\rightarrow$ D'  $\rightarrow$ D. This information flow process involves some normal in-network propagations and the \emph{choices} of  some users that the information is worth propagating also in the other network. It is worth noticing that this model should clarify and support Jenkins' critical view in using a viral model to describe flows of information --- as soon as we depart from a very general and simple model the active role of users becomes more evident.

The right hand side of Figure~\ref{reachability} shows that even for users that are already connected to each other in one or both networks their distance may decrease. Also in this case this depends on the transfer of information from one network to the other.

\begin{figure}
\begin{center}
\includegraphics[width=.70\columnwidth ]{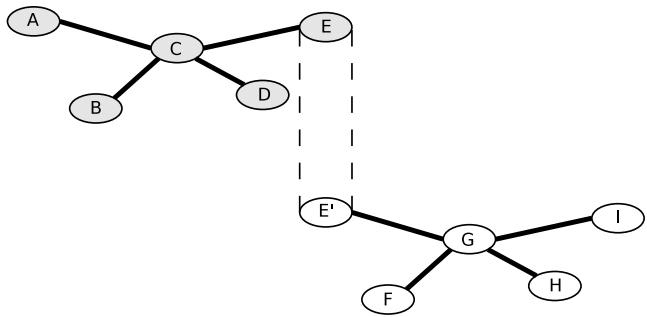} 
\end{center}
\caption{A user (E) marginal to both his/her ego-networks but with high betweenness in the Multi Stratum model}
\label{betweenness}
\end{figure}

Finally, consider Figure~\ref{betweenness}. Here we show how a user that can be marginal (with low betweenness) in all his/her single networks may become central when we consider his/her Multi Stratum network. In this case this happens because user E exploits the two networks to keep in touch with different communities, and as such he/she can act as a bridge between them passing information from one community to the other.

From the point of view of computing closeness and betweenness centrality, both measures are based on the concept of distance between two nodes: closeness measures the average inverse distance of one node from the others and betweenness counts the number of shortest paths passing through a node. The previous examples highlight how we can compute the extended distances by considering a single network obtained starting from the MLN:

\begin{definition}[flat]
Let $\mbox{MSN} = (L_1, \dots, L_n, \mbox{IM})$ be a Multi Stratum Network. flat(MSN) = $(V,E)$ where:
\begin{itemize}
\item $V = \bigcup_i V_i$
\item $E  = \bigcup_i E_i \cup \bigcup_{i,j} \mbox{IM}_{i,j}$ 
\end{itemize}
\end{definition}

\begin{definition}[distance]
The distance $d$ of two nodes $u, v$ in a MSN is defined as:
\[ 
d(u, v) = \mbox{min}\{ d(u', v') \ | \ u' \in [u]_{eq}, v' \in [v]_{eq}, u', v' \in flat(\mbox{MSN}) \}
\]
\end{definition}

\section{Experimental results}\label{experiments}

Our experimental phase has been designed to investigate how the use of the MSM impacts on our ability to describe a complex real world situation made of users with multiple accounts on several social network sites. Therefore we have selected a group of users each one with an active account on three well known social network sites: FriendFeed, Twitter and YouTube. Two of these services -- Friendfeed and Twitter --  appear to be very similar regarding their general goal: both are microblogging services (even if Friendfeed is more an aggregator of on-line content) and both allow the sharing of various kinds of information toward a list of followers. In both cases there is no technical requirement of reciprocity in the following/follower relationship. On the opposite YouTube, while also based on a Social Network structure, is a web service dedicated to video content sharing. YouTube offers different ways to connect the users to each others: users may be connected through a \emph{friendship} relationship -- where reciprocity is required or through a \emph{following} connection -- where reciprocity is not required. According to a Multi-layered approach \cite{DBLP:journals/tsmc/KazienkoMK11,DBLP:conf/socialcom/KazienkoBMG10} we could also establish a connection when a user comments on or interacts with a video uploaded by another user or the two users interact with the same video content uploaded by a third user. 

In the following analysis we focus on YouTube's \emph{friendship} stratum and Twitter and Friendfeed \emph{following} strata.

\subsection{Data extraction}

Friendfeed is a social media aggregator. In this system while users can directly post messages and comment on other messages much like in Facebook and other similar SNSs, they can also register their accounts on other systems. In this way, using the Friendfeed API we could retrieve the multiple accounts of the same users for several social services.

In our dataset, which is available for download on the project Web site (http://larica.uniurb.it/sigsna) we collected 7 628 users who registered exactly one Twitter account and exactly one YouTube account and whose Twitter and YouTube accounts were associated to exactly one Friendfeed user, this to remove collective accounts not corresponding to single persons and to build a Pillar model.  The final step of the data extraction phase consisted in the retrieval of the followers/friends of these users on each network (October 2011). Due to privacy settings and to the dinamicity of the networks this information was not available for all users in our sample, therefore we deleted users for which the information was not available.

As a result we obtained a Friendfeed network with 7 677 120 arcs, a Twitter network with 37 805 211 arcs and a YouTube network with 708 911 arcs. These networks were used for the analysis of degree centrality correlations.
In addition, we also built three networks by keeping only those connections between users in our sample. The three networks, respectively with 37 997, 67 123 and 1 185 arcs, were used for the error estimation of degree centrality and for the tests on closeness/betweenness.

\subsection{Data analysis}

A first analysis of the collected data is aimed at showing how different users behave differently on different SNSs and thus how a wider perspective, such as the MSM, is required. As a starting point for our analysis we decided to focus on the measurement of the degree centrality on the three networks we are dealing with. While the degree centrality shows many limits as a pure centrality measure \cite{Freeman1979a} it is undoubtedly a powerful measure to describe the overall presence of a single user within a social network and a good indicator of his/her potential use of the network itself. While, in fact, we cannot assume that a large value of degree corresponds to a current intensive use of the network by the users, the social nature of these connections -- that have to be directly established even when they do not require any reciprocity -- suggest an explicit effort made by the users in terms of network presence. We are in fact well aware that these connections might be no longer in use (users often move from a SNS to a newer one in order to follow their group of peers \cite{BoydDanahGolderScott2010})  but these might be used as an effective indicator of the amount of activity that the users put into a specific SNS.  

\begin{figure}
\begin{center}
\includegraphics[ angle=-90, width=.47\textwidth ]{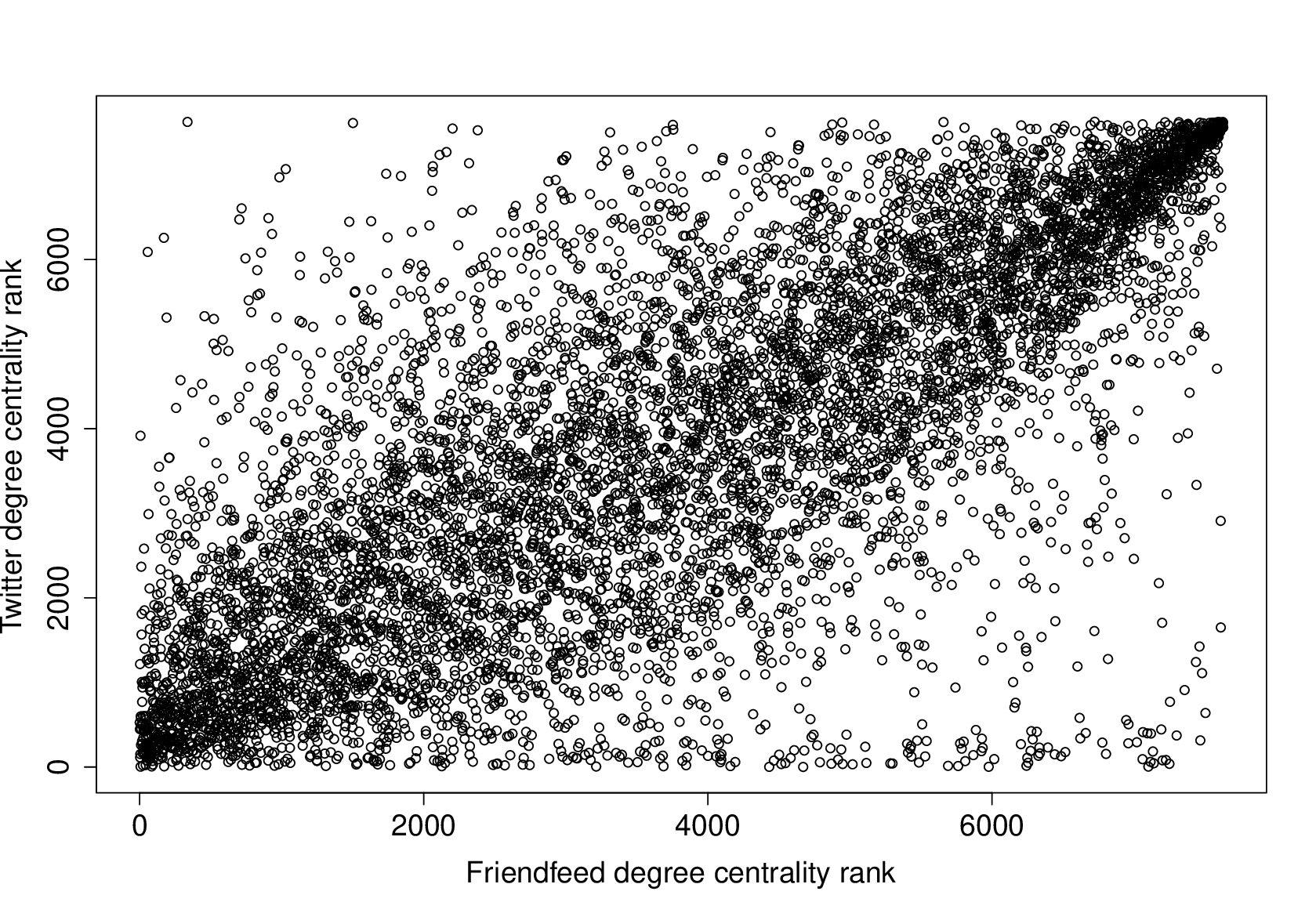}
\includegraphics[ angle=-90, width=.47\textwidth ]{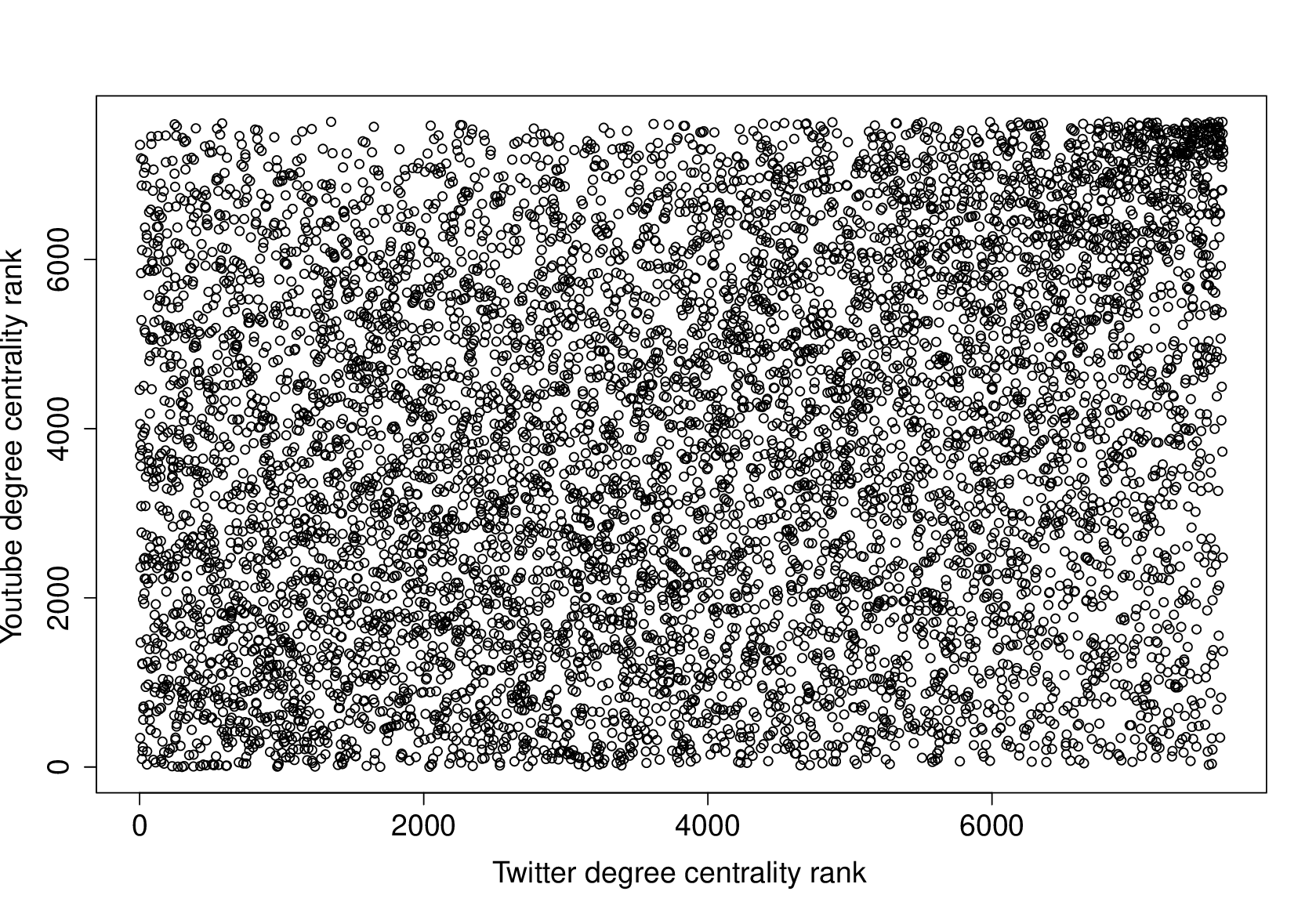} 
\end{center}
\caption{User ranking (according to their degree centrality) in different networks: Friendfeed and Twitter (left) and Twitter and YouTube (right)}
\label{fig:degree_centrality}
\end{figure}

The left hand side of Figure~\ref{fig:degree_centrality} shows the correlation between user rankings according to their degree centrality on the Twitter network and on the Friendfeed network, while on the right of  Figure~\ref{fig:degree_centrality} we have shown the correlation between user rankings according to their degree centrality on the Twitter network and on the YouTube network. To interpret these figures consider that each point represents a user, and users with a high x or y coordinate are among the top users on the corresponding SNS according to their degree centrality (more precisely, x and y coordinates correspond to the ranking of the user, 0 for the user with lower degree centrality, up to 7 628 for the user with the highest degree centrality in that SNS).

The comparison between the two correlations shows an interesting phenomenon. The visualization of the correlation between Friendfeed and Twitter shows an overall linearity that means that users with a high degree centrality on Twitter seem to maintain it on the Friendfeed network. On the opposite when we compare the degree centrality ranking on Twitter with the YouTube network we are unable to detect any clear kind of linearity, what we see instead is a very dispersed scenario with little correlation between the two rankings. Following this visualization, we can compute a \emph{Network Complementarity Index} (NCI) defined as the Pearson correlation between user rankings in two networks. In our data, NCI(Friendfeed, Twitter) = 0.75, while NCI(YouTube, Twitter) = 0.21.

This result requires a deeper discussion in order to be fully explained. As we have claimed before Friendfeed and Twitter show an interesting similarity according to the goal of the platform. Both can be used to share short text messages or images with a group of contacts.  On the opposite YouTube is generally perceived as a completely different online service that allows users to share their video content with both a list of friends and a broader public. This difference in the social use (and perception) of the two platforms is crucial to be able to understand properly the data we have shown. Despite the fact that all the three systems can be described as SNSs, user behaviour within them changes according to their social perception \cite{DBLP:conf/asonam/MagnaniR11}. Users, in other words, are active players in their online social experience and they constantly move through very different situations. A single user might have a very active account on Twitter while his YouTube account has been created just to be able to comment on a couple of videos but it has never been used for any social networking activity. In addition to this our data suggest that the similarity in user's behaviour within the platform is related with the purpose of the platform itself. Therefore the NCI we have introduced earlier allows us to define the similarity of two networks from a user behaviour perspective.

While the NCI allows us to compare different user behaviours within different networks from a \emph{social} perspective these data also mean that, when it comes to the analysis of online user activity, taking into account only a single system might provide data largely misleading and different from the real global online experience.

\begin{figure}
\begin{center}
\includegraphics[ angle=-90, width=.42\textwidth ]{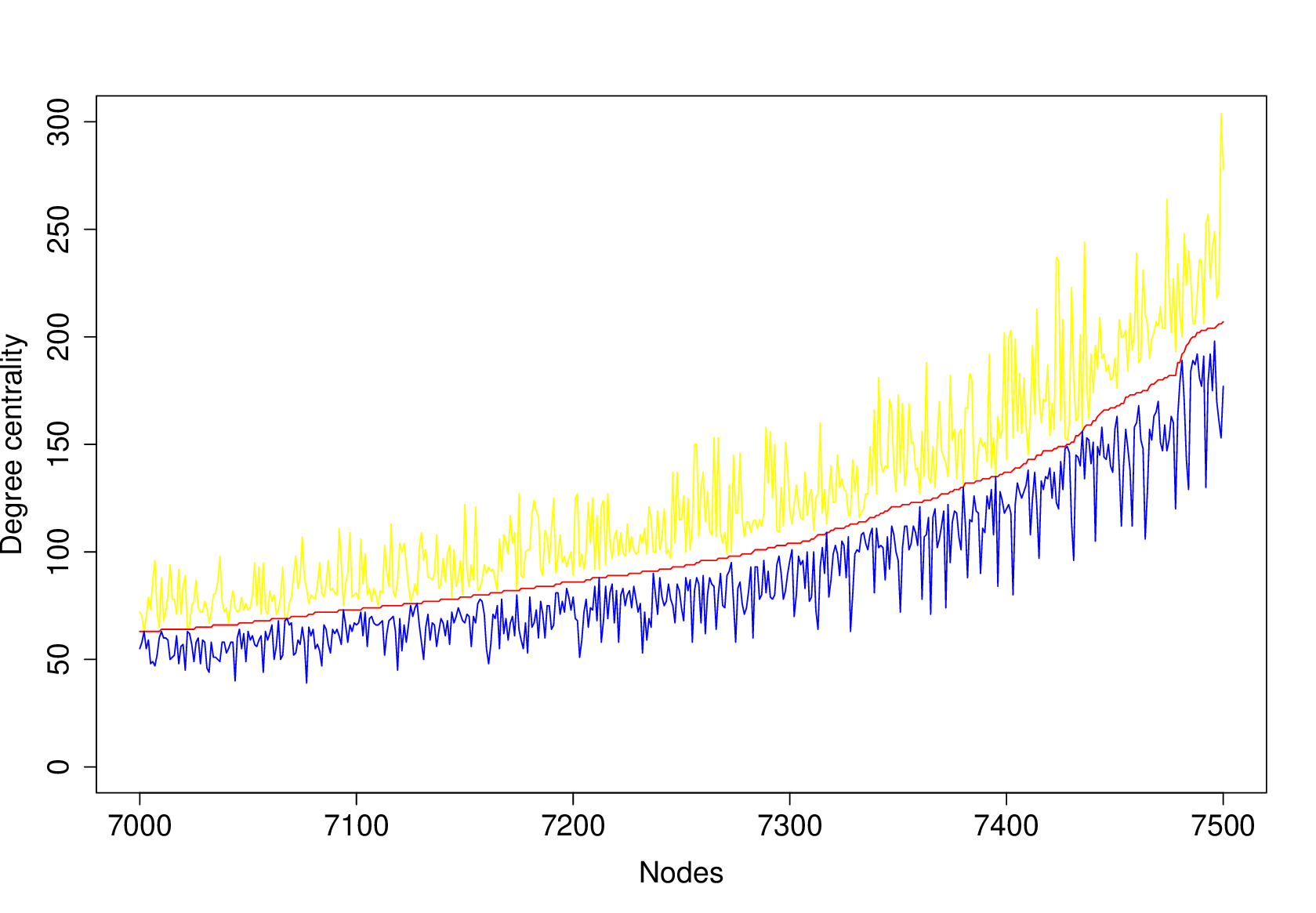}
\includegraphics[ angle=-90, width=.42\textwidth ]{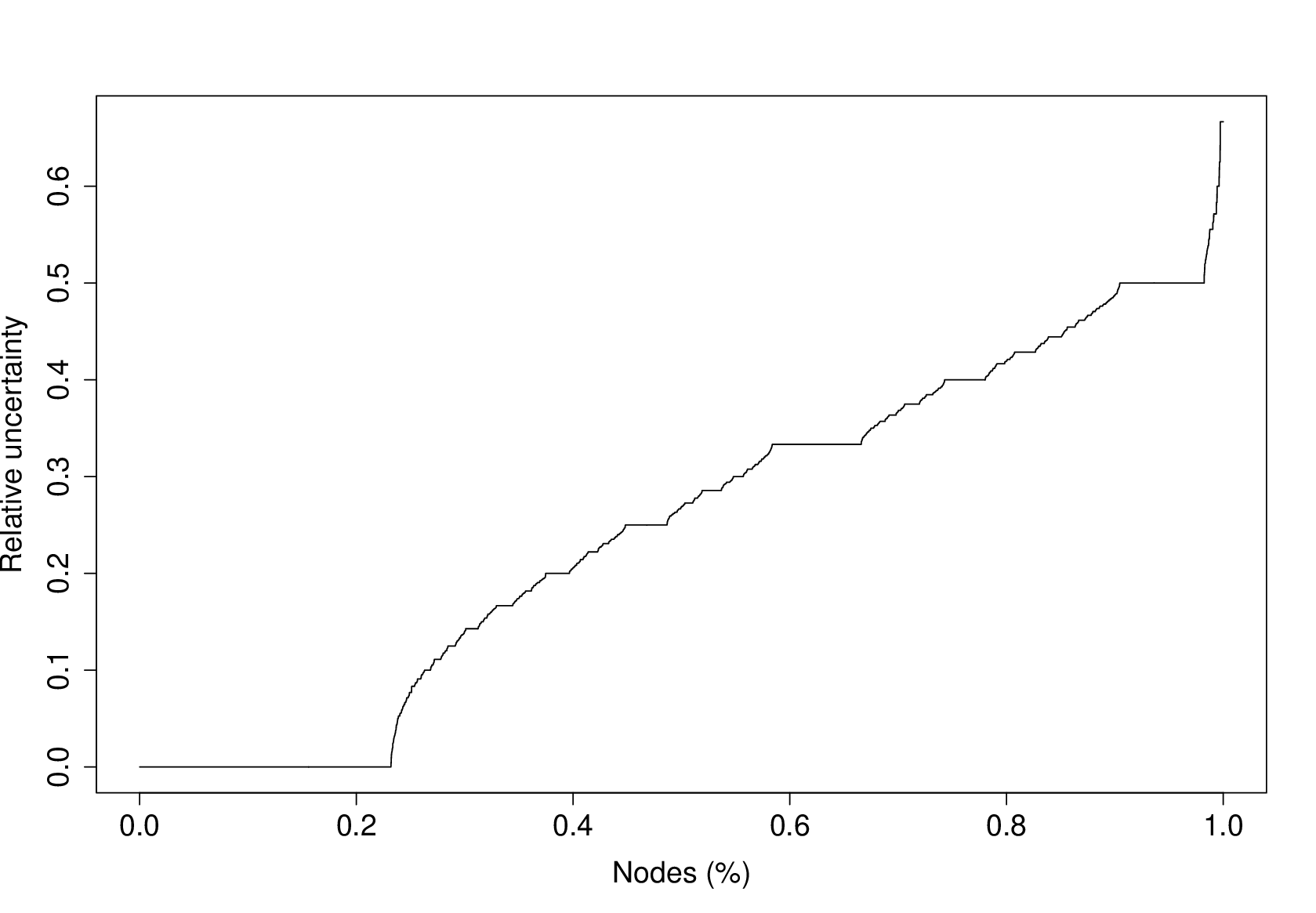} 
\end{center}
\caption{Multi Stratum Degree Centrality, and the absolute (left) and relative (right) error that we can experience using the single degree centralities in all the distinct networks}
\label{fig:degree_error}
\end{figure}

In order to be able to define how large this error can be we performed the analysis represented in Figure~\ref{fig:degree_error}. Assume a user has three strata and 100 connections in each stratum: what is his/her Multi Stratum degree centrality? In the worst case, he/she could be connected to the same 100 people on all networks, with an overall degree centrality of 100. In the best case, all connections in the three networks would be distinct, leading to an overall degree centrality of 300. In general, if we indicate with $\delta_i$ a user's degree centrality on his/her $i^{th}$ stratum, his/her Multi Stratum degree would be in the interval $[\min_i(\delta_i), \sum_i \delta_i]$. This can also be regarded as the absolute level of uncertainty with respect to a Multi Stratum degree centrality when we only consider the single values of degree on the single networks.

The first graph (Figure~\ref{fig:degree_error}) shows the minimum value (bottom line), the maximum value (upper line) and the value on the MSN of the degree centrality (central line) within the three networks, for a sample of 500 nodes sorted by their Multi Stratum Degree Centrality. The distance between the two boundary lines shows the actual error range described above computed on the real data, which is -- de facto -- a level of uncertainty on the specific measure. Therefore, the subsequent question naturally arising regards how much we could be wrong in estimating the degree centrality of a user from the set of his/her single layer degree centralities.

The second graph in  Figure~\ref{fig:degree_error} shows the relative uncertainty on the degree centrality values over the percentage of nodes with that level of uncertainty. For more than 20\% of the nodes the potential error in the estimation is $0$ --- this happens for all users active on at most one network. However, this graph shows how only less than 40\% of users have a uncertainty value lower than .2, i.e., while estimating their Multi Stratum degree centrality we could be 20\% wrong. For about 10\% of the users (right-most users in the x axis) the potential error in the estimation would be around 50\%. This is clearly a huge and unsustainable level of uncertainty for a very important and basic measure like degree centrality.

The evaluation of the uncertainty range supports our major claim that online social activities present such a high level of variability that any approach aiming at describing them by flattening on a single dimension will introduce an unsustainable level of error. These errors can be mitigated by the adoption of a model such as the MSM.

We conclude this experimental analysis on the usage of the MSM by discussing its impact on the distance between nodes, that is at the basis of the computation of centrality measures such as closeness centrality and betweenness centrality. First, we can notice that as expected the number of reachable nodes increases significantly when these values are computed on the MSN. The Giant component in the Friendfeed network connects 3 159 nodes and the Giant component in the Twitter network includes 4 990 nodes, while in the MSN this number increases to 5 512, about $10\%$ more than in the most connected of the single networks composing it. This increment shows how the adoption of a MSN provides a larger perspective by including nodes that are -- de facto -- connected but that would be ignored by a traditional approach.


The second experimental evidence regards the distance among nodes. In Table~\ref{avgdist} we show a sample of ten nodes indicating how the average distance between them \emph{and all the other nodes already reachable through the Friendfeed and Twitter networks} decreases on the MSN. When nodes are observed within a MSM perspective they have a larger number of viable paths that can be used and often, as it emerges from Table~\ref{avgdist}, the possibility to move from a \emph{stratum} to another produces a reduction of the distance between two nodes. This phenomenon can be easily understood by thinking of our everyday experience: if we are directly connected with \emph{user A} on YouTube we will use that network to communicate with him/her instead of using Twitter where \emph{user A} is, as an example, two nodes away from us. 

\begin{table}[htdp]
\caption{Average distance of a random sample of ten individuals computed on Friendfeed (FF), Twitter (TW) and the MSN}
\begin{center}
\begin{tabular}{|l|l|l|l|}
\hline
 & Avg FF distance & Avg TW distance & Avg MSN distance \\
\hline\hline
user \#1 & 5.15 &4.04 &3.98 \\
\hline
user \#2 & 4.72 &3.51 &3.46 \\
\hline
user \#3 & 4.19 &4.08 &3.66 \\
\hline
user \#4 & 3.85 &3.27 &3.14 \\
\hline
user \#5 & 3.91 &3.38 &3.19 \\
\hline
user \#6 & 3.33 &3.27 &2.96 \\
\hline
user \#7 & 0.0 &0.0&0.0\\
\hline
user \#8 & 0.5 &1.0&0.5\\
\hline
user \#9 & 4.21 &4.04 &3.70 \\
\hline
user \#10 & 3.72 &2.98 &2.88 \\
\hline
\end{tabular}
\end{center}
\label{avgdist}
\end{table}%

\section{Conclusion}
In this paper we have proposed a model for the representation of multi stratum networks, together with extensions of classical SNA centrality measures. Based on this proposal, other centrality and SNA metrics will be extended in the future. We argued that MSNs constitute an important element toward a more complex approach to SNA applied to real-world online social activities. The application of these metrics to a real multi stratum network has confirmed our claims by showing how considering a complete network-of-networks model allows us to extract results from our analyses that do not correspond completely to the ones that can be obtained from each network separately. A MSN approach gives us a more complete picture allowing us to have a better perception of the real complexity hidden behind every online social activity. We like to think at this new approach as a sort of Gestaltic Social Network Analysis where the full experience of the online activity is grater than the sum of its parts \cite{Sternberg2009}. This kind of approach that takes into consideration the whole online presence of the user produces different and more accurate descriptions of the centrality and the connections available to the single on-line user opening new research perspectives on many fields of online SNA.

In particular, we think that introducing multi stratum models in different kinds of studies of on-line networks could boost open research directions and potentially open new ones, and in particular the study of the connections between different strata, i.e., the dynamics of information propagation from one network to the others, and also the topic of data integration for social networks that appears to be a fundamental activity to build unified user profiles from distributed on-line accounts.

Finally, we have also introduced several new concepts and analytical tools that we consider a valuable addition to the current available network analysis toolbox, and in particular: the \emph{Multi-Stratum network}, the \emph{Pillar model}, \emph{Node Mappings}, \emph{Extended Centrality metrics}, the concept of \emph{Gestaltic Network Analysis} and the \emph{Network Complementarity Index}.



\end{document}